\documentstyle[12pt,epsf,aps]{revtex}  

\begin{document}
\newcommand{\p}{\partial}
\newcommand{\ls}{\left(}
\newcommand{\rs}{\right)}
\newcommand{\beq}{\begin{equation}}
\newcommand{\eeq}{\end{equation}}
\newcommand{\beqa}{\begin{eqnarray}}
\newcommand{\eeqa}{\end{eqnarray}}
\newcommand{\bdm}{\begin{displaymath}}
\newcommand{\edm}{\end{displaymath}}
\draft
\title{
Influence of the in-medium pion dispersion relation 
in heavy ion collisions
}
\author{
C. Fuchs, L. Sehn, E. Lehmann, J. Zipprich and 
Amand Faessler
}
\address{
Institut f\"ur Theoretische Physik der Universit\"at T\"ubingen,\\
Auf der Morgenstelle 14, D-72076 T\"ubingen, Germany
}
\maketitle  
\begin{abstract}
We investigate the influence of medium corrections to the 
pion dispersion relation on the pion dynamics in 
intermediate energy heavy ion collisions. To do so a 
pion potential is extracted from the in-medium 
dispersion relation and used in QMD calculations and thus 
we take care of both, real and imaginary part of the 
pion optical potential. The potentials are determined from different 
sources, i.e. from the $\Delta$--hole model and from phenomenological 
approaches. Depending on the strength of the potential 
a reduction of the anti-correlation of pion and nucleon 
flow in non-central collisions is observed as well as an 
enhancement of the high energetic yield in transverse pion spectra. 
A comparison to experiments, in particular to $p_t$-spectra 
for the reaction Ca+Ca at 1 GeV/nucleon and the pion in-plane flow in 
Ne+Pb collisions at 800 MeV/nucleon, generally favours a weak potential. 
\vspace{5mm}
\end{abstract}
\section{Introduction}
The main motivation to study heavy-ion reactions at intermediate
energies is to extract some information on nuclear matter under
extreme conditions, i.e. at high densities and/or high temperatures. 
Besides nucleonic observables like the rapidity distribution 
and flow observables as, e.g., 
the bounce-off and squeeze-out also mesons \cite{metag93,ehehalt93} 
($\pi, K^+, \eta$ etc.) emitted from the reaction zone
can be probes of excited nuclear matter. 
Pion yields and spectra 
\cite{harris87,brill93,venema93,schwalb94,muentz95,gillitzer96} 
turned out not to be sensitive to properties of the nucleon-nucleon 
interaction and difficult to interpret. 
Due to their strong interaction with the nuclear environment 
pionic observables at the freeze out are the result of complex 
creation and rescattering processes. However, pions can provide 
information about the resonance production in compressed and 
excited nuclear matter, in particular with respect to the 
$\Delta$(1232) resonance which is the dominant production channel. 
In high energetic collisions with heavy nuclei the creation of 
"resonance matter", i.e. a state with a density of $\Delta$-resonances 
comparable to the nuclear saturation density has, e.g., been discussed 
in this context \cite{metag93,ehehalt93}. 

Whereas total pion yields and low energetic $p_t$-spectra can be well 
explained by present theoretical transport approaches \cite{Ba95} the 
subthreshold production of high energetic pions is a question of current 
debate. Such spectra have, e.g., been measured by the TAPS 
\cite{venema93,schwalb94} and the KaoS \cite{muentz95} collaborations and 
at the Fragment Separator \cite{gillitzer96} at GSI. Various processes 
can contribute to the creation of subthreshold pions. One is the 
accumulation of energy by multistep collision processes, e.g. 
$\Delta \Delta$-scattering. But also heavier resonances as the 
$N^*$(1440) can be excited and decay into high energy pions. 
The analysis in particular of these high energy 
$p_t$-spectra is supposed to yield information about the role 
of resonances in heavy ion collisions \cite{schwalb94}. 
However, both cases are rare and require 
a sufficiently high density of excited resonances, a state which mainly 
occures in heavy systems. In lighter systems which have also been measured 
(Ar+Ca by TAPS \cite{schwalb94} and Ni+Ni in Ref. \cite{gillitzer96}) 
multistep processes and the creation of higher resonances are less 
probable and it is not clear if these phenomena are sufficient to 
explain the experimental data. 

However, most theoretical approaches \cite{Ba95,Li91} 
included the interaction of the pions
with the surrounding nuclear medium only by collision processes, i.e.
parametrizing the imaginary part of the pion optical potential. 
In the present work we include also the real
part of the pion optical potential which influences the pion 
propagation through the nuclear medium. We thus take care of the 
full in-medium pion optical potential. 
We investigate the influence on 
$p_t$-spectra with pion momenta of several hundred MeV/c. This 
allows to test the in-medium dispersion relation in the high energy range. 

To investigate the low energy range of the dispersion relation 
we consider the momenta of the pions in the collective in-plane pion flow 
which have values of some 10 MeV/c. The pion flow 
as been measured, e.g., by the DIOGENE group for the system Ne+Pb at 
800 MeV/nucleon \cite{diogene}. The prominent observation was there an 
always positive pion flow in contrast to the typical S-shape of the 
nucleon flow. This fact is probably due to the projectile-target 
geometry in this highly asymmetric system and the reabsorption of pions 
by the target matter, i.e. the so called shadowing effect. 
In symmetric systems this shadowing effect may even lead to a complete 
anti-correlation of pion and nucleon flow \cite{Ba95,zip96}. 
Although conventional calculations, i.e. without 
a pion potential, are able to qualitatively reproduce the DIOGENE data 
\cite{Li91,hartnack88} in particular this (anti)correlation of flow reacts 
sensitive on the in-medium effects taken into account in the present 
work.

The knowledge of the pion 
optical potential from elastic pion--nucleus scattering is, however, 
restricted to nuclear densities at and below saturation and relatively 
small energies \cite{SCM79}. In intermediate energy heavy ion 
reactions up to 2 GeV incident energy per nucleon 
baryon densities of three times saturation density are reached and 
transverse momenta of about 1 GeV/c are detected in 
pion spectra. In this range the 
real part of the pion potential is almost unknown. Thus, 
one has to extrapolate the pion dispersion 
relation to the ranges relevant for heavy ion collisions.
As done in Ref. \cite{zip96} we apply two models, i.e. the 
$\Delta$--hole model \cite{We88,Fr81} and a phenomenological ansatz 
suggested by Gale and Kapusta \cite{Ga87}. The application 
of these models allows to investigate the influence of the 
possible boundary cases, i.e. a soft 
($\Delta$--hole) and a rather strong phenomenological 
potential. Moreover, since the in-medium corrected dispersion 
relation suggested by Gale and Kapusta is close to the hypothetic pion 
condensation limit it is a question of interest 
whether the boundary conditions occuring in intermediate 
energy heavy ion collisions would allow such a scenario. 
In addition we apply a third potential which is a modification 
of the parametrization of Gale and Kapusta and lies in strength 
between the two former cases. In Ref. \cite{zip96} we have demonstrated 
the influence on pionic flow observables in heavy ion collisions. Here 
we want to derive quantitative statements and to distinguish between 
the different approaches by comparison to experimental data, i.e. 
to $p_t$-spectra of the emitted pions and the pion in-plane 
transverse flow produced in non-central collisions.

The paper is organized as follows: First we give an overview of the 
treatment of pions in the Quantum Molecular Dynamics (QMD) approach, 
in particular with respect to the resonances included and the collision 
processes. Next we discuss the dispersion relation and the resulting 
pion potentials and then compare the results to experiment. 
Finally a summary and an outlook is given.

\section{Pions in Quantum Molecular Dynamics}

QMD is a semiclassical transport model which accounts for relevant 
quantum aspects like the Fermi motion of the nucleons, stochastic 
scattering processes including Pauli blocking in the final states, 
the creation and reabsorption of resonances
and the particle production. A detailed description 
of the QMD approach can be found in Refs. \cite{Ai91,khoa92}.

Each baryon is represented by a Gaussian wave packet
with a fixed width $L$. The temporal evolution of the centroids of 
these wave packets
is described by the classical equations of motion generated from an $N$-body 
Hamiltonian 
\beqa
\frac{\p {\bf p}_i}{\p t} = - \frac{\p H_B }{\p {\bf q}_i} \quad ,  \quad
\frac{\p {\bf q}_i}{\p t} =   \frac{\p H_B }{\p {\bf p}_i} 
\label{motion1}
\eeqa
with 
\beq
H_B = \sum_i \sqrt{{\bf p}_{i}^2 + M_{i}^2} + 
\frac{1}{2}\sum_{i,j\atop (j\neq i)} \left( U_{ij} + U_{ij}^{Yuk} 
+ U_{ij}^{coul}\right)
\quad .
\label{hamB}
\eeq
The Hamiltonian, Eq. (\ref{hamB}), contains mutual two- (and three-) 
body potential interactions which are finally determined as 
classical expectation values
from local Skyrme forces $ U_{ij}$ supplemented by a phenomenological
momentum dependence and an effective Coulomb interaction $U_{ij}^{coul}$ 
\beqa
 U_{ij} &=& \alpha\left(\frac{\rho_{ij}}{\rho_0}\right)
+\beta\left(\frac{\rho_{ij}}{\rho_0}\right)^{\gamma}
+\delta ln^2 \left(\epsilon |{\bf p}_i -{\bf p}_j |^2 +1 \right)
\frac{\rho_{ij}}{\rho_0}
\label{skyrme} 
\\
 U_{ij}^{coul} &=& \left( \frac{Z}{A}\right)^2 
\frac{e^2}{|{\bf q}_i- {\bf q}_j |} 
erf\left(\frac{|{\bf q}_i-{\bf q}_j|}{\sqrt{4L}}\right)
\label{coul}
\eeqa 
where $\rho_{ij}$ is a two-body interaction density defined as 
\beq
\rho_{ij} = \frac{1}{(4\pi L)^{\frac{3}{2}}}
e^{-( {\bf q}_{i}- {\bf q}_{j})^2 /4L}
\label{dens1}
\eeq
and $erf$ is the error function. The parameters $\alpha, \beta, \gamma, 
\delta, \epsilon$ of the Skyrme interaction, Eq. (\ref{skyrme}), 
are determined in order to
reproduce simultanously the saturation density 
($\rho_0 = 0.16 fm^{-3}$) and the binding energy ($E_B$=-16 MeV) for
normal nuclear matter for a given incompressibility as well as the correct
momentum dependence of the real part of the 
nucleon-nucleus optical potential \cite{khoa92,Ai86}. In the present 
calculations we apply the standard parametrization of a hard/soft 
equation of state (K=380/200 MeV). The Yukawa-type potential 
$ U_{ij}^{Yuk}$ in Eq. (\ref{hamB}) mainly serves to improve 
the surface properties and the stability of the initialized nuclei. 

Analogously to the baryons, Eq. (\ref{hamB}), also the mesons 
obey Hamilton's equations of motion. The Hamiltonian of the pions, i.e. 
the sum of the respective single particle energies 
$\omega_i$, is represented in a mean field form 
\beqa
H_\pi = \sum_{i}^{N_\pi} \omega_i 
      = \sum_{i}^{N_\pi} \left( \sqrt{{\bf p}_{i}^2 + m_{\pi}^2} + 
Re V^{opt}_\pi ({\bf p}_{i},{\bf q}_{i}) \right)
\label{hamPi}
\eeqa
where $N_\pi$ is the actual number of pions. 
The dependence of the pionic mean field, i.e. the 
real part of the pion optical potential $ Re V^{opt}_\pi$, 
on the pion coordinates originates from the medium effects 
in dense nuclear matter. Such medium effects are naturally 
expressed through a density dependence and $ Re V^{opt}_\pi$ takes the form 
\beq
Re V^{opt}_\pi ({\bf p}_{i},{\bf q}_{i}) = 
 Re V^{opt}_\pi \left({\bf p}_{i},\rho_B ({\bf q}_{i}) \right)
\label{piopt}
\eeq 
where $\rho_B$ is the respective baryon density. In conventionell 
approaches \cite{Li91,Ba95} the Hamiltonian, Eq. (\ref{piopt}), 
is taken as that 
of the vacuum, i.e. $ Re V^{opt}_\pi$ is set equal zero.

Hard core scattering of the particles is included by the simulation 
of the collision processes by standard Monte Carlo procedures. 
The collisions probabilities are determined by a geometrical 
minimal distance criterium $d\leq\sqrt{\sigma_{tot}/\pi}$ weighted 
by the Pauli blocking factors of the final states \cite{Ai91,bg88}. 
For the inelastic 
nucleon-nucleon channels we include the $\Delta(1232)$ as well as 
the $N^{*}(1440)$ resonance. In the intermediate energy range the 
resonance production is dominated by the $\Delta$, however, the 
$N^{*}$ yet gives non-negligible contributions to the high energetic 
pion yield \cite{metag93}. The 
resonances as well as the pions originating from their decay are 
explicitely treated, i.e. in a non-perturbative way and all relevant 
channels are taken into account. 
In particular we include the resonance production and rescattering 
by inelastic NN collisions, the one-pion decay of $\Delta$ and 
$N^{*}$ and the two-pion decay of the $N^{*}$ and one-pion 
reabsorption processes. These are summarized in Table 1 
where the isospin dependence of the various channels is suppressed 
for the sake of simplicity; which is, however, taken into account 
in the calculations. 

For the cross sections of the inelastic NN channels 
we adopt the parametrizations of Ref. \cite{Hu94} 
which have been determined from one-boson-exchange amplitudes 
in Born approximation. The lifetimes of the resonances are 
determined through their energy and momentum dependent decay widths 
\beq
\Gamma (|{\bf p}|) = \frac{a_1 |{\bf p}|^3}
{(1+ a_2 |{\bf p}|^2 )(a_3 + |{\bf p}|^2) } \Gamma_0
\label{reswidth}
\eeq
which originates from the $p$-wave representation of the 
resonances. In Eq. (\ref{reswidth}) ${\bf p}$ is the momentum of the 
created pion (in GeV/c) in the resonance rest frame. 
According to Ref. \cite{Hu94} the values 
$a_1$=22.83 (28.8), $a_2$=39.7 and $a_3$=0.04 (0.09) are used for 
the $\Delta$ ($N^*$) and the bare decay widths are taken as  
$\Gamma_{0}^\Delta$= 120 MeV and $\Gamma_{0}^{N^*}$= 200 MeV. 
The reabsorption cross sections ($\pi N\rightarrow \Delta,N^{*}$) 
are determined assuming a Breit-Wigner distribution for the masses, i.e. 
\beq
\sigma = \frac{\sigma_0}{{\bf p}^2 (M_R \Gamma_R )^2}
\frac{1}{ (s-M_{R}^2 )^2 + (M_R \Gamma_R )^2 }
\label{sigres}
\eeq
where $R$ stands for a $\Delta$ or $N^*$ and ${\bf p}$ is the c.m. 
momentum. In contrast to, e.g., Ref. \cite{Li91} the momentum 
dependent resonance width, Eq. (\ref{reswidth}) is used in 
Eq. (\ref{sigres}). The resonance rescattering 
cross section ($N(\Delta,N^{*})\rightarrow NN$) is determined 
from detailed balance. To take care of the proper phase space 
available for this channel we take the finite width 
of the resonances into account 
as proposed in Ref. \cite{DanBe91}. In particular for pions 
near the $\Delta$-threshold this procedure leads to a 
significant enhancement of the rescattering cross section and 
the low energetic pion yield is suppressed by about 30\%. 
Thus we are able to reasonably 
reproduce the total pion multiplicities with results similar to 
Ref. \cite{Ba95}. 

In the presence of a pion potential this has to 
be taken into account in the energy balance 
in order to ensure energy conservation in 
the collision processes, i.e. the relation 
\beq
\sqrt{ {\bf p}_{R}^2 + M_{R}^2 } = 
\sqrt{ ({\bf p}_{R}-{\bf p}_{\pi})^2 + M_{N}^2 } +
\sqrt{ {\bf p}_{\pi}^2 + m_{\pi}^2 } + 
Re V^{opt}_\pi \left({\bf p}_{\pi},\rho_B \right)
\label{conserv}
\eeq
has to be fulfilled. Concerning the pion reabsorption 
Eq. (\ref{conserv}) is exactly fulfilled; in the case of a resonance 
decay this procedure is more involved since the decay takes place in 
the resonance rest frame. The potential of the created pion depends, however, 
on its momentum in the rest frame of the colliding nuclei. Thus the 
pion momentum is determined iteratively until Eq. (\ref{conserv}) 
is fulfilled with an accuracy better than 0.5 MeV which corresponds to 
an energy violation of less than about 0.04\%. 

Although the elementary cross sections \cite{Hu94} are parametrizations of the 
free cross sections medium corrections to the imaginary part of the pion 
optical potential are included firstly by the enhancement of the 
rescattering/reabsorption probability which is proportional to 
the nuclear density, i.e. the available scattering partners, and 
secondly by the Pauli blocking in the final states 
according to the respective phase space occupancy. Via Eq. (\ref{conserv}) 
also the real part of the pion potential gives corrections 
to the imaginary part. Thus the mean 
free path of the pions is significantly reduced and a consistent 
treatment of real and imaginary part of the optical potential 
is achieved. Furthermore, the momentum dependence of 
the nuclear mean field, Eq. (\ref{skyrme}), 
results in a (non-relativistic) effective mass of the baryons. 

\section{The pion potential}
In contrast to the imaginary part of the pion optical potential
which is well known from inelastic pion-nucleus scattering 
the knowledge of the real part extracted from elastic pion-nucleus 
scattering is rare and restricted to low energies \cite{SCM79}. 
Data are available only for momenta up to about 2$m_\pi$c. However, 
in intermediate energy heavy ion reactions pion momenta of more than 
1 GeV/c can occure and the high energy tails of the pion spectra are 
of particular interest since they are supposed to yield information 
about the resonance production in hot and dense nuclear matter 
and corresponding in-medium effects \cite{venema93,schwalb94,gillitzer96}. 
Furthermore, nuclear densities 
up to three times saturation density can be reached whereas the 
elastic pion-nucleus scattering is restricted to densities 
around saturation and below. Thus, one has to extrapolate the dispersion 
relation to the energy range relevant for heavy ion collisions. 
However, such an extrapolation is affected by large uncertainities. To cover 
the unknown ranges we apply two contrary approaches, i.e. a microscopic 
and a phenomenological ansatz for the pion dispersion relation.

The microscopic ansatz is based on the perturbation expansion of the 
$\Delta$-hole model \cite{We88}. The summation of the $\Delta$-hole 
polarization graphs results in a pion self energy $\Pi$ 
entering into the in-medium pion dispersion relation
\begin{eqnarray}
\omega({\bf p})^2= {\bf p}^2+m_{\pi}^2+\Pi(\omega,{\bf p},\rho_B)
\quad .
\label{disp1}
\end{eqnarray}
Here the self energy depends on the energy $\omega$ and the 
momentum ${\bf p}$ of the pion and on the nuclear matter density $\rho_B$. 
In the framework of the $\Delta$-hole model the self energy is 
finally given in the form
\begin{eqnarray}
\Pi(\omega,{\bf p},\rho_B)=\frac{{\bf p}^2 
\chi(\omega,{\bf p},\rho_B)}{1-g'\chi(\omega,{\bf p},\rho_B)}
\label{self1}
\end{eqnarray}       
with
\begin{eqnarray}
\chi(\omega,{\bf p},\rho_B)&=&-\frac{8}{9}
\left(\frac{f_{\Delta}}{m_{\pi}}\right)^2 
\frac{\omega_{\Delta}({\bf p}) \rho_B}
{\omega_{\Delta}^2({\bf p})-\omega^2({\bf p})}
\nonumber\\
\omega_{\Delta}&=&\sqrt{M_{\Delta}^2+{\bf p}^2}-M_N
\quad .
\nonumber
\end{eqnarray}
The parameters entering into Eq. (\ref{self1}), in particular 
the $\pi N\Delta$ coupling constant $f_{\Delta}$ 
and the correlation parameter $g'$ are taken in consistence 
with the OBE parameters of Ref. \cite{Hu94} and a consistent 
treatment of the real and imaginary part of the pion optical 
potential is achieved.

\begin{figure}
\begin{center}
\leavevmode
\epsfxsize = 10cm
\epsffile[30 85 430 710]{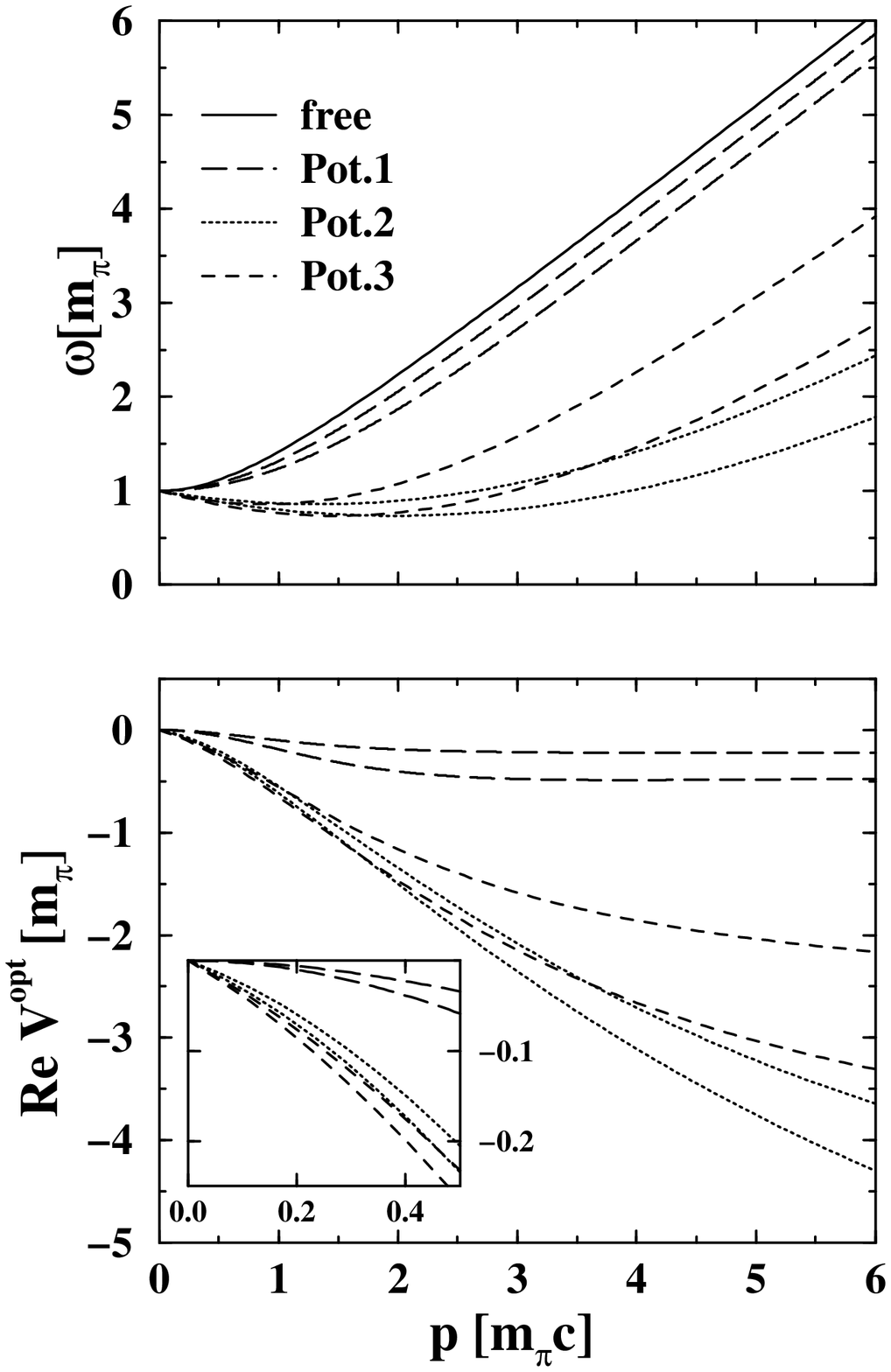}
\end{center}
\caption{
Upper Figure: The free pion dispersion relation (solid) is compared to 
various in-medium dispersion relations at nuclear densities 
$\rho_B = \rho_0$ (respective upper curves) and $\rho_B = 2\rho_0$ 
(respective lower curves). The in-medium dispersion relation 
is extracted from the $\Delta$-hole model (Pot.1, long-dashed) 
and the phenomenological parametrization of Ref. 
\protect\cite{Ga87} (Pot.2, dotted) as well a softer 
parametrization (Pot.3, dashed) are shown. 
Lower Figure: Real part of the pion optical potential 
at densities $\rho_B = \rho_0$ (upper curves) and $\rho_B = 2\rho_0$ 
(lower curves) extracted from the corresponding 
dispersion relations. The inserted figure zoomes the low momentum 
range.
}
\label{fig1}
\end{figure}

Although this approach works 
reasonably well at low baryon densities and low pion momenta \cite{Fr81} 
one has to be careful when appying Eq. (\ref{self1}) to heavy ion 
reactions. The self energy obtained from the $\Delta$-hole model 
includes beside of excitation of $\Delta N^{-1}$ states 
also short range correlations of these states. In this 
approximation one neglects, however, terms of 
higher order \cite{Di87} which are necessary to prevent 
the system to undergo a phase transition to the so called pion 
condensation. Furthermore, Eq. (\ref{disp1}) yields 
the wrong boundary conditions for high energy pions when 
they pass through the surface of nuclear matter into the vacuum, 
in particular a "quasi-pion" expressed by a 
$\Delta$-hole state has to convert asymptotically to a real 
pion which can be detected.  
A common practice to avoid these unphysical features 
is to mix the two solutions of the dispersion relation, Eq. (\ref{disp1}), 
i.e. the pion-like and the $\Delta$-hole-like branch ($\Omega_1,\Omega_2$) 
in order to obtain an effective pion dispersion relation \cite{Eh93}
\beqa
\omega_{eff} = Z_1  \Omega_1 +  Z_2  \Omega_2 \quad ,\quad
 Z_1 +  Z_2 =1
\quad .
\eeqa
The probabilities for the quasi-pion to sit on the 
pion-like branch ($ Z_1$) or the $\Delta$-hole-like branch ($ Z_2$) 
are determined from the condition
\beq
\frac{1}{\omega^2 - {\bf p}^2 - m_{\pi}^2 - \Pi} =
\frac{Z_1}{ \omega^2 - \Omega_{1}^2} + 
\frac{Z_2}{ \omega^2 - \Omega_{2}^2}
\eeq
and thus the physical boundary conditions 
are fulfilled. Replacing $\omega$ in Eq. (\ref{disp1}) 
by $\omega_{eff}$ one 
obtaines an effective pion self energy $\Pi_{eff}$. The real part 
of the optical potential is obtained from the pion wave equation, i.e. 
the Klein-Gordon equation as \cite{We88}
\beq
Re V^{opt}_\pi \left({\bf p},\rho_B \right) =
\frac{\Pi_{eff}(\omega_{eff},{\bf p},\rho_B )}{2\omega_{eff}} 
\quad .
\label{vopt1}
\eeq
The density and momentum dependence of the dispersion relation 
relation are shown in Fig. \ref{fig1} and it is seen that such a construction 
leads to a strong softening of the in-medium effects compared, e.g., 
to the original dispersion relation \cite{Eh93}. Consequently, 
the medium dependence of the resulting potential, Eq. (\ref{vopt1}), 
(called Pot.1 in the following and shown in Fig. \ref{fig1}) is 
moderate and the attraction of the potential is weak. Hence one can't be 
sure that the modified dispersion relation still represents the 
true pion-nucleon interaction.

As already done in Ref. \cite{zip96} we also consider a 
phenomenological ansatz suggested by Gale and Kapusta 
\cite{Ga87}. The dispersion relation then reads 
\begin{eqnarray}
\omega(p)&=& \sqrt{(|{\bf p}|-p_0)^2+m_0^2} - U 
\label{disp2}
\\
U &=&\sqrt{p_0^2+m_0^2}-m_{\pi}
\label{disp3}
\\
m_0&=&m_{\pi}+6.5(1-x^{10})m_{\pi}
\label{disp4}
\\
p_0^2&=&(1-x)^2m_{\pi}^2+2m_0m_{\pi}(1-x) 
\quad .
\end{eqnarray}
A phenomenological medium dependence is introduced via 
$x=e^{-a(\rho_B /\rho_0)}$ with the parameter $a=0.154$ and 
$\rho_0$ the saturation density in nuclear matter 
(here $\rho_0 = 0.16 fm^{-3}$). 
The form of Eqs. (\ref{disp2}-\ref{disp4}) is motivated 
by the following constraints \cite{Ga87}:
\begin{enumerate}
\item The group velocity $\p \omega/\p p$ may not 
exceed the velocity of light. 
\item For high energetic pions many-body effects should be of 
minor importance and $\omega$ should in leading order 
be proportional to $p$ for $p\to \infty$. 
\item Corresponding to a strong p-wave interaction the energy 
should have a minimum.
\item Due to a weak s-wave interaction medium effects are 
weak for pions at rest with respect to the surrounding medium.  
\item The observation of pionic atoms implies that 
$\omega \simeq $-20 MeV for $\rho_B  = \rho_0$ and 
$p \simeq 2 m_\pi c$. 
\item Pion condensation can only appear at infinite density. 
\end{enumerate}
The dispersion relation as well as the corresponding potential 
(called Pot.2 in the following) are shown in Fig. \ref{fig1}. 
Here $Re V^{opt}_\pi$ is extracted from Eq. (\ref{disp2}) 
as $Re V^{opt}_\pi = \omega - \sqrt{{\bf p}^2 +m_{\pi}^2}$. 
It turns out that the potential is much stronger and its medium 
dependence is much more pronounced than in the case of the 
modified $\Delta$-hole dispersion relation. In this context we want 
to mention that this effect is to a large extent due to the mixing of 
the pionlike and the $\Delta$-hole-like branches in Pot.1 since a 
potential extacted according Eq. (\ref{disp2}) from the pure 
pionlike branch given in $\Delta$-hole model is of 
the same magnitude as Pot.2 \cite{xiong93}. 
Thus, Pot.1 and Pot.2 can be looked as the boundary cases of 
a rather weak and a very strong pion potential. Furthermore, 
the phenomenological dispersion relation 2 is closer to the 
limit where pion condensation can occur. 

In order to test the pion dispersion relation it is therefore a 
natural step to suggest a third parametrization (Pot.3) which lies in 
between the former ones. To do so, we decrease the power $x^{10}$ 
in Eq. (\ref{disp4}) to $x^{2}$. The constraintes (i)-(iv) 
remain unrendered by this modification except of point (v) 
where now a value of +10 MeV is obtained instead of -15 MeV in the former 
case. However, we do not consider this as a severe drawback 
since the application of the potential to heavy ion collisions 
should not be affected therefrom. In addition, also 
the $\Delta$-hole model potential does not yield a bound quasi-pion for 
these values of $\rho_B$ and $p$. 

\section{Results}
\begin{figure}
\begin{center}
\leavevmode
\epsfxsize = 15cm
\epsffile[0 85 430 430]{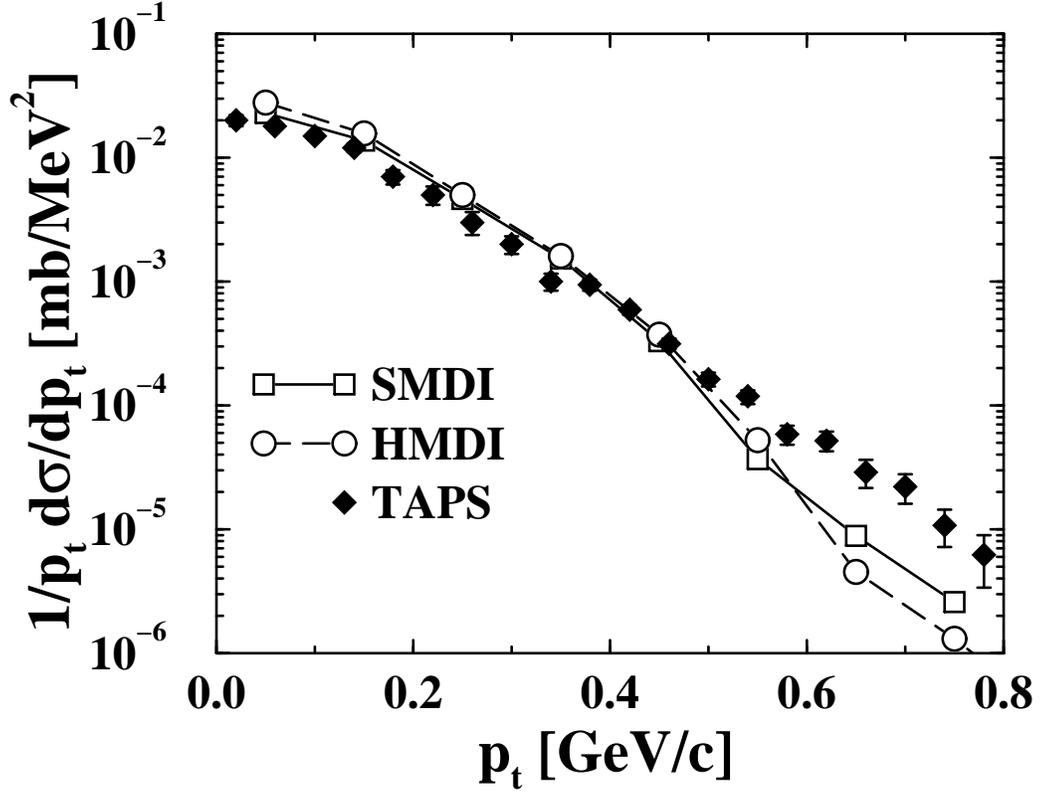}
\end{center}
\caption{
Influence of the nuclear EOS on the $\pi^0$ $p_t$-spectrum for the 
reaction $^{40}Ca$+$^{40}Ca$ at 1 GeV/nucleon. A hard (HMDI) and a soft 
(SMDI) momentum dependent Skyrme force have been used. The experimental 
data heave been measured by the TAPS Collaboration for the system 
$^{40}Ar$+$^{40}Ca$ at 1 GeV/nucleon \protect\cite{schwalb94}.
}
\label{fig2}
\end{figure}

\begin{figure}
\begin{center}
\leavevmode
\epsfxsize = 15cm
\epsffile[0 85 430 430]{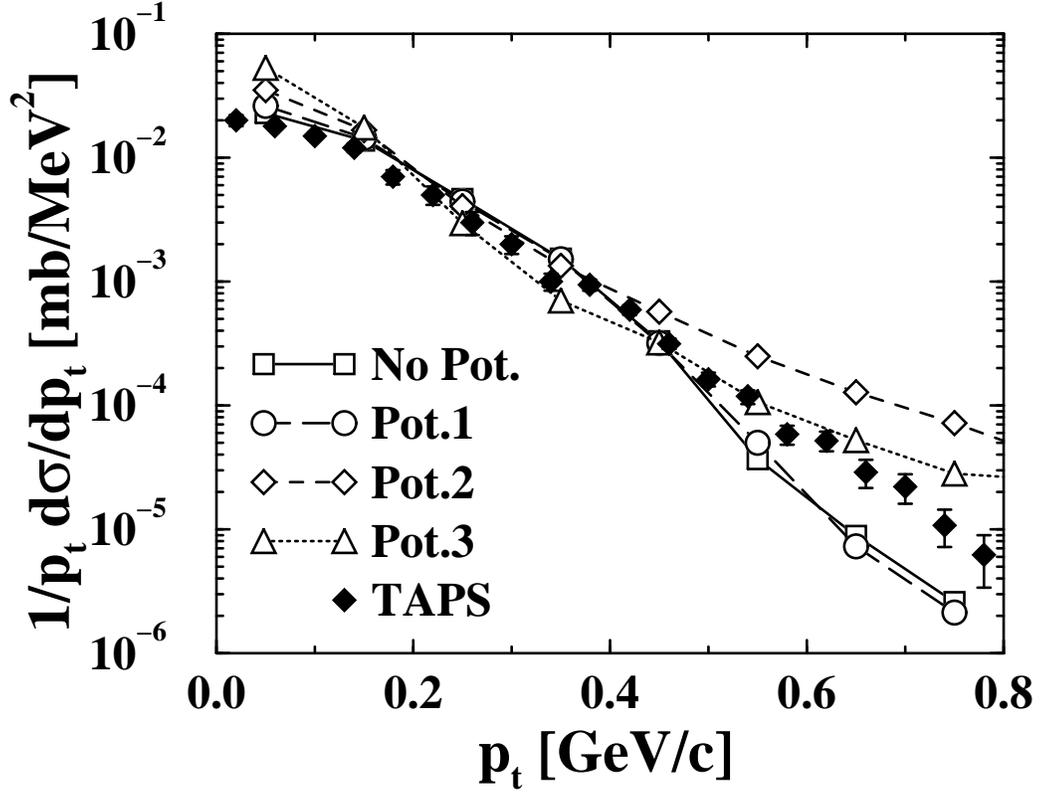}
\end{center}
\caption{
Influence of the in-medium pion potential on the $\pi^0$ $p_t$-spectrum 
for the same reaction as in fig.2. The calculations are performed 
without (solid) and including a pion potential. The respective potentials 
have been obtained from the $\Delta$-hole model (Pot.1) and by the 
phenomenological ansatz of Ref. \protect\cite{Ga87} (Pot.2). 
Pot.3 corresponds to a modification of Pot.2. 
}
\label{fig3}
\end{figure}

First we consider transverse momentum $p_t$-spectra for the system 
$^{40}Ca$+$^{40}Ca$ at 1 GeV incident energy per nucleon. In particular 
we compare to $\pi^0$-spectra recently measured by the TAPS 
collaboration \cite{schwalb94}. Here the $\pi^0$'s have the advantage 
that they are not distorted by Coulomb effects. The calculations are 
impact parameter averaged and a rapidity cut of $-0.2\leq y_{c.m.}\leq 0.2$ 
has been applied which takes into account the detector acceptance. 
The data shown in Figs. \ref{fig2} and \ref{fig3} have been 
obtained with this cut 
\cite{averbeck96} and are slightly enhanced with respect to the results 
shown in Ref. \cite{schwalb94} where a larger cut has been used. 
In fig.2 we first investigate the influence of the nuclear equation 
of state and therefore compare a soft (SMDI) and a hard (HMDI) 
momentum dependent Skyrme force, see Eq. (\ref{skyrme}). It is seen 
that the dependence of the pion spectrum on the nuclear EOS is 
moderate and the agreement with the data is reasonable, however, not 
overwhelmingly good. We want to mention that the results obtained with 
the hard EOS are close to those of Ref. \cite{Ba95}. As also found 
there the low $p_t$-spectrum is slightly overestimated whereas the 
high $p_t$ range is underestimated by about a factor of 3. 
In Ref. \cite{Ba95} the lack of high energy pions is even more pronounced 
than in the present calculations. This is probably due to the $N^*$(1440) 
resonance which was not taken into account there.

Next we turn to the influence of the in-medium pion potential. In Fig. 
\ref{fig3} we show the $\pi^0$-spectra for the same 
reaction as in Fig. \ref{fig2} now including 
the pion-nucleon interaction as discussed in the previous section. 
All calculations are performed with the soft momentum dependent 
Skyrme force. It is seen that the weak $\Delta$-hole potential (Pot.1) 
has nearly no influence on the $p_t$-spectrum which is in agreement with 
the results of Ref. \cite{Eh93} where a similar potential has been 
investigated. In this context we want to mention that concerning 
the low energy pions the present results (and those of Ref. \cite{Ba95}) 
stand somehow in contradiction to the BUU calculations of Refs. 
\cite{Eh93,xiong93} since there a significantly underestimated low energy 
$p_t$ yields was obtained when the pion potential was neglected. 
However, in these works comparisons to other measurements 
have been performed \cite{berg91,bevelac}. In our opinion the 
discrepancies are rather caused by the analysis of the data 
than by a slightly different treatment 
of the pion creation and annihilation processes outlined in section 2.  

In the case of the phenomenological potentials 2 and 3 
the situation is completely different now. With Pot.2 the 
low $p_t$-range is slightly 
enhanced, however, the high energy pions are overestimated 
by nearly one order of magnitude. Such an enhancement of low energy 
pions has been observed in Ref. \cite{xiong93} where an effective 
pion potential has also 
be determined within the $\Delta$-hole model which is, however, apparently 
stronger than Pot.1 in the present work. This complex 
behavior can be understood 
by the strong attraction of the respective potential which 
forces the pions to follow the trajectories of the nucleons. 
Most pions get bound by the stopped participant matter resulting in 
an enhancement of the low $p_t$-yield. On the other handside, pions 
which are bound by the spectator matter are driven out to high 
transverse momenta by the nucleonic flow and thus the high $p_t$-range 
is strongly enhanced. This effect is diminished when the 
potential is weaker (Pot.3). Here the very high energy 
yield ($p_t \ge 0.7$GeV/c) 
is still slightly overestimated but also the low $p_t$-range 
is now overestimated by about a factor of 3. The latter is 
undestandable since for low momenta ($p\leq 1.5 m_\pi c$) Pot.3 is even 
stronger than Pot.2, see zoomed region in Fig. \ref{fig1}.

\begin{figure}
\begin{center}
\leavevmode
\epsfxsize = 15cm
\epsffile[0 85 430 380]{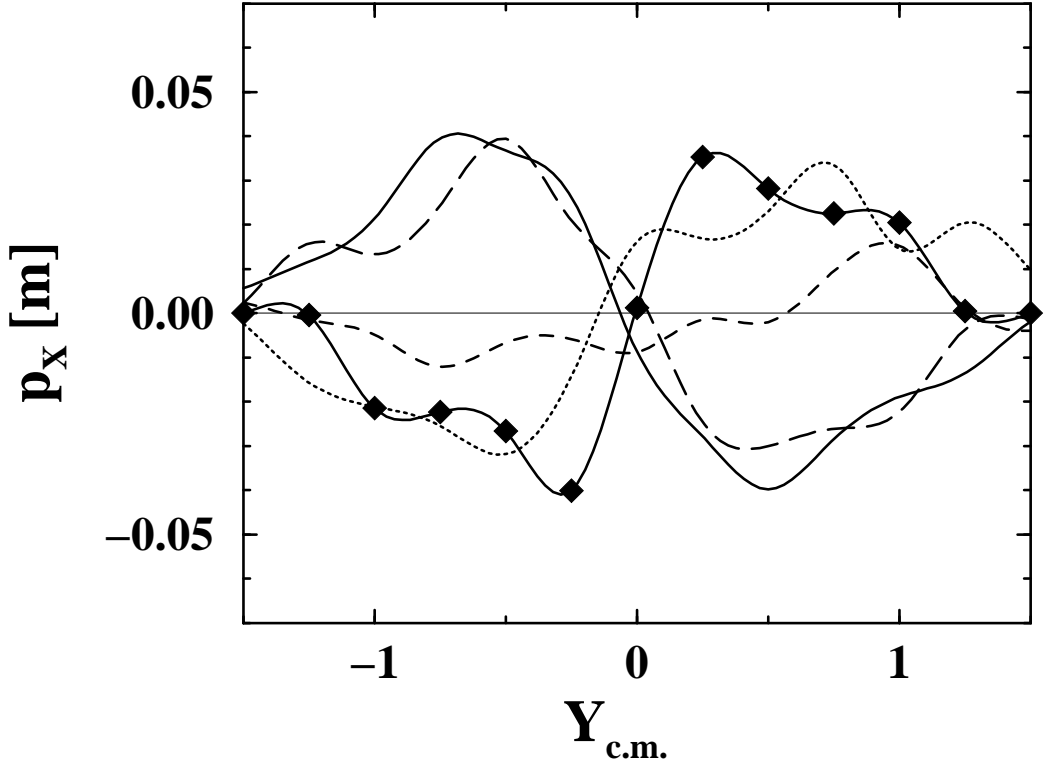}
\end{center}
\caption{
Influence of the in-medium pion potential on the 
in-plane tansverse pion flow as a function of the center-of-mass 
rapidity for the same reaction as in fig.2. 
The calculations are performed 
without (solid) and with inclusion of a pion potential. 
The respective potentials have been obtained from the 
$\Delta$-hole model (Pot.1, long-dashed) and by the 
phenomenological ansatz of Ref. \protect\cite{Ga87} (Pot.2, dotted). 
Pot.3 (dashed) corresponds to a modification of Pot.2. 
In addition the nucleon flow is shown (solid with diamonds). 
The results are scaled to the pion and nucleon mass 
and thus the transverse flow is given in units of $m_\pi c$ and 
$m_N c$, respectively.
}
\label{fig4}
\end{figure}

The above observations are clearly reflected in the pion 
transverse flow. In Fig. \ref{fig4} we compare the pion and 
the baryon in-plane 
transverse flow for the reaction Ca+Ca at 1 GeV/nucleon. The results are 
impact parameter averaged and scaled to the pion and nucleon mass, 
respectively. By this scaling pion and nucleon flow are of the same 
order of magnitude. It is seen that without any pion-nucleon interaction 
the flow of pions and nucleons is clearly anti-correlated. The 
anti-correlation originates from the absorption of the pions by the 
participant matter which produces the nucleonic flow. This is known as 
the shadowing effect. In the presence of the weak $\Delta$-hole 
potential (Pot.1) the pion flow remains nearly the same. 
In the case of Pot.2 the pion flow switches 
from an anti- to a correlation with the nucleon flow, i.e. the pions 
are forced to follows the trajectories of the spectator nucleons 
by the strong attractive potential. 
Further it is seen that the strength of Pot.3 has just a magnitude where 
the attraction and the shadowing effect are nearly completely 
counterbalanced and the resulting flow is around zero. 
This overall behavior is in good agreement with the interpretation of 
$p_t$-spectra given above. 

In summary none of the cases under consideration is able to 
reproduce the experimental spectrum over the entire range of energy 
with a satisfying high accuracy. Both, the low as well as the high 
$p_t$-range react sensitive on the pion potential. Pot.2 seems to be 
too attractive and can be ruled out by the comparison to the data. 
However, the present results 
indicate that the inclusion of higher resonances may be not sufficient to 
explain the role of high energy pions but furthergoing medium effects 
should be taken into account.

Next we investigate the collective in-plane transverse flow of pions 
which has been measured for the system Ne+Pb at 800 MeV/nucleon by the 
DIOGENE collaboration \cite{diogene}. There the most striking results was 
the observation of a positive pion flow also for backwards rapidities, 
i.e. the pion flow is partially anti-correlated to the nucleon flow. 
In particular here we consider a semi-central reaction at impact parameter 
b=3 fm. For a comparison with the data we included the experimental detector 
filter cuts given in Ref. \cite{diogene}. In addition, we simulated the 
reconstruction of the estimated reaction plane as it was done in the 
experimental analysis. In contrast to the theoretical calculation where 
the true reaction plane is known a priori in the experiment the transverse 
in-plane vector is estimated for every event by
\beq
{\bf Q}= \sum_j \omega_j {\bf p}_{\perp j}
\eeq
where the sum runs over all detected protons weighted by their 
relative rapidity $\omega_j = y_i - <y>$ with $<y>$ the mean rapidity 
of the total system. The estimated in-plan pion momentum 
$p_x =  {\bf p}_{\perp} \cdot {\hat{\bf Q}}$ is then obtained by the 
projection on the unit vector ${\hat{\bf Q}}$.

\begin{figure}
\begin{center}
\leavevmode
\epsfxsize = 15cm
\epsffile[0 85 430 380]{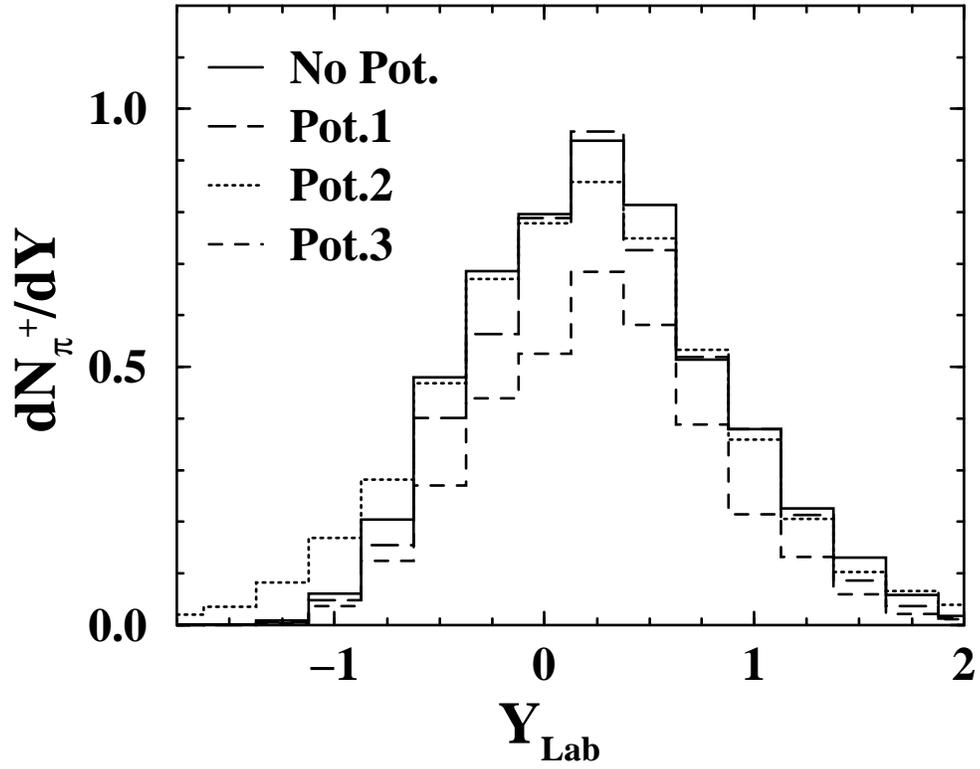}
\end{center}
\caption{
$\pi^+$ rapidity distribution for a semicentral (b=3 fm) Ne on Pb 
reaction at 800 MeV/nucleon after applying the experimental detector 
filter cut. The calculations are performed 
without (solid) and including a pion potential. 
The respective potentials 
have been obtained from the $\Delta$-hole model (Pot.1, long-dashed) 
and by the 
phenomenological ansatz of Ref. \protect\cite{Ga87} (Pot.2, dotted). 
Pot.3 corresponds to a modification of Pot.2.
}
\label{fig5}
\end{figure}

Fig. \ref{fig5} shows the corresponding $\pi^+$ rapidity distribution 
obtained with the various pion potentials and including 
the detector cuts. 
The calculation without pion potential is thereby in a good agreement 
with the result of Ref. \cite{Li91}. The influence of the pion potentials 
1 and 2 on the longitudinal flow is relatively weak, only in the 
case of Pot.2 
a slight enhancement of the backward, i.e. the target-like, 
rapidity distribution is observed. 
Applying Pot.3 we observ a strong suppression of the detected 
$\pi^+$ yield over the entire rapidity range. This effect is also 
reflected in the total yields given in Table 2; is, however, not so 
pronounced since the total yield is only suppressed by about 10\%. 
In generel the usage of a potential leads to a slight reduction of the 
total yields which was also found in Refs. \cite{Eh93,xiong93}. 
Concerning Pot.3 the effect seen in Fig. \ref{fig5} 
seems to be somehow an artefact of the detector geometry. 
Nevertheless, the dynamics of the pions is significantly changed 
and slow pions near the $\Delta$-threshold get in particular strongly bound 
and and are preferrentially reabsorbed by the spectator matter.

\begin{figure}
\begin{center}
\leavevmode
\epsfxsize = 15cm
\epsffile[0 85 430 380]{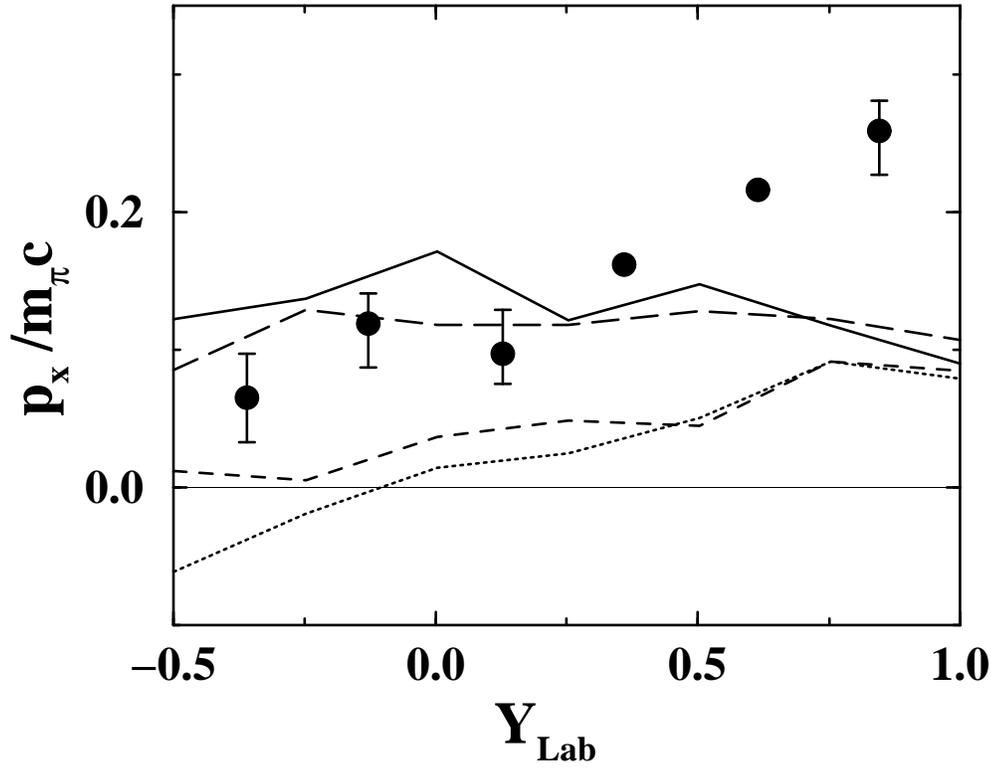}
\end{center}
\caption{
Transverse in-plane flow per pion sacled to the pion mass 
for a semicentral (b=3 fm) Ne on Pb 
reaction at 800 MeV/nucleon after applying the experimental detector 
filter cut. The calculations are performed 
without (solid) and including a pion potential. 
The same potentials as in fig. 5 have been used, i.e. 
Pot.1 (long-dashed), Pot.2 (dotted) and Pot.3 (dashed). The 
data are taken from Ref. \protect\cite{diogene}. 
}
\label{fig6}
\end{figure}

In Fig. \ref{fig6} the corresponding in-plane flow 
per pion is shown in units of 
the pion mass. Here the procedure to reconstruct the experimental 
reaction plane has been performed. It turns out that in the case 
of a vanishing or a weak potential (Pot.1) the flow is always positive 
and in a fairly good agreement with the data for values around 
mid-rapidity. In the high forward rapidity range we, 
however, underestimate the data by about a factor of two. Since the 
results are strongly distorted by the large asymmetry of the 
considered system and the reconstruction of the reaction plane 
this effect can be due to a lack of sufficiently 
good statistics in this range. The positive flow at backward 
rapidities is explained by the reabsorption of the pions 
by the large target and thus is a consequence of the 
shadowing effect. In the case of the strong Pot.2 the shadowing 
effect is overcompensated by the attraction of the potential 
and the flow shows a definit 
change of sign from negative to positive values at midrapidity. 
Similar as in Fig. \ref{fig4} Pot.3 shows the same behavior 
which is, however, not so pronounced. Both, 
Pot.2 and 3 yield significantly too less positive flow over the 
entire rapidity range.

These observations are also 
reflected in the ratio $R$ between the numbers of pions with positive and 
negative values of $p_x$ displayed in Table. 2. Since we did not explicitely 
include the Coulomb interaction in the propagation of the pions, Eqs. 
(\ref{motion1}) and (\ref{hamPi}), the different charge states 
are only taken into account via the isospin dependence of the respective 
creation/absorption channels. Thus, for a fair comparison to the 
measured ratio a mean value (corrected for the respective total yields) 
of $\pi^+$ ($R=1.42$) and $\pi^-$ ($R=1.30$) 
which is about $R=1.34$ should be compared. Then the theoretical 
positive $p_x$ abundancy is a reasonable agreement with the experiment 
for the calculations without pot. and with Pot.1. 
In the case of Pot.2 and Pot.3, however, R is close to unity.

\section{Summary and Conclusions}
We investigated the influence of medium corrections to the pion 
dispersion relation on the pion dynamics in heavy ion collisions. 
This was done by the introduction of a pion-nucleus potential 
through which the pions propagate between their collisions and 
refers to the real part of the pion optical potential. The imaginary part 
is included by the non-perturbative description of elementary 
rescattering and reabsorption processes and is directly 
medium corrected via the respective densities of nucleons and resonances. 
Hence a consistent treatment of the real and imaginary part of the 
optical potential was achieved. 

The pion potential was extracted from the dispersion relation given in 
the $\Delta$-hole model thereby mixing the pion-like and the 
$\Delta$-hole-like branches in order to avoid some unphysical 
features of the model. This, however, leads to a strong softening of the 
medium effects. Thus we also applied a phenomenological 
dispersion relation with a pronounced medium dependence. 

Studying the influence of such effects 
the different approaches have been subjected to a comparison to the 
experiment for two contrary observables, i.e. spectra and in-plane flow. 
Thereby one has to keep in mind that 
pions at the freeze out are remanents of a complex collision history, i.e. 
of multiple creation and reabsorption processes. The analysis of the 
DIOGENE flow data implies a preferrential abroption of pions by the 
spectator matter known as the shadowing effect which may even lead to an 
anti-correlation of the collective pion and nucleon in-plane flow. 
However, a strong attractive pion potential leads to a bending of the pions 
by the nucleons and thus was found to favour a correlation of the 
respective flow. The magnitude of this effect is directly proportional to 
the strength of the potential. Since the pions get bound by the participant 
matter (slow pions) as well as by the spectator matter (fast pions) 
this process is complex, i.e. an enhancement of the low $p_t$-range as well 
as of the high $p_t$-range was observed in transverse momentum spectra. 

A comparison to experiments, i.e. the TAPS spectra and to the DIOGENE 
flow data, seems to rule out a too attractive optical potential. 
Hence the pions created in intermediate energy heavy ion collisions 
are far from the limit of pion condensation. 
The weak effective $\Delta$-hole potential has nearly no influence on 
pionic observables as well as the dependence on the nuclear equation of state 
was found to be rather weak. However, in particular the spectra of 
high energetic pions react sensitiv on the dynamics and the present 
results would favour a pion potential which's attraction is weak at 
low energies but becomes more pronounced with increasing energy. 
In our opinion a furthergoing analysis of high energetic subthreshold 
spectra as, e.g., measured in Ref. \cite{gillitzer96} may help to 
clarify the question if conventional approaches as, e.g., energy accumulating 
multi-scattering processes or the creation of higher resonaces are 
sufficient to explain such data. The present results 
indicate that the inclusion of in-medium corrections to the 
real part of the pion optical potential is of essential importance for 
a correct description of pion dynamics. In addition the analysis of the 
pion in-plane flow also in symmetric systems as, e.g. 
recently measured by the FOPI 
collaboration, will help to learn something about the 
pion dispersion relation from heavy ion collisions.



\newpage
\newpage
\begin{table}
\begin{tabular}{cc}
 $NN\leftrightarrow N\Delta$      & $\Delta\leftrightarrow N\pi$ \\
 $NN\leftrightarrow N N^* $       & $N^*   \leftrightarrow N\pi$  \\
 $NN\rightarrow \Delta\Delta$     & $N^*   \rightarrow \Delta\pi$ \\
 $N\Delta\rightarrow \Delta\Delta$& $N^*   \rightarrow N\pi\pi$ \\
\end{tabular}
\caption{
Inelastic scattering processes which are included in the 
present calculations.
}
\label{restab}
\end{table}
\begin{table}
\begin{tabular}{ccccrc}
            & No Pot.  & Pot.1  & Pot.2 & Pot.3 & Exp. \\ \hline
$N_{\pi^+}$ & 1.89     & 1.86   & 1.80  & 1.59  & \\
$R$         & 1.26     & 1.25   & 1.03  & 1.12  & 1.34 \\
\end{tabular}
\caption{
Total $\pi^+$ yield (without experimental filter) and the ratio $R$ 
of detected $\pi^+$ and $\pi^-$ pions with positive to negative values of 
$p_x$ for the reaction Ne on Pb at 800 MeV/nucleon and impact parameter 
b=3 fm. The experimental value for $R$ is an average of the respective 
values for $\pi^-$ and $\pi^+$ given in Ref. \protect\cite{diogene}.
}
\label{ratio}
\end{table}

\end{document}